\documentstyle[aps,manuscript]{revtex}



\begin{document}

\title{Fractal escapes in Newtonian and relativistic multipole gravitational fields }

\author{Alessandro P. S. de Moura\thanks{
email: sandro@ifi.unicamp.br
}}

\address{Instituto de F\'{\i}sica Gleb Wataghin, UNICAMP, 13083-970 Campinas SP, Brazil}

\author{Patricio S. Letelier\thanks{
email: letelier@ime.unicamp.br
}}

\address{Instituto de Matem\'{a}tica, Estat\'{\i}stica e Ci\^{e}ncia da Computa\c{c}\~{a}o,
Departamento de Matem\'{a}tica Aplicada, UNICAMP, 13083-9790 Campinas SP, Brazil}

\maketitle
\begin{abstract}
We study the planar motion of test particles in gravitational fields produced
by an external material halo, of the type found in many astrophysical systems,
such as elliptical galaxies and globular clusters. Both the Newtonian and the
general-relativistic dynamics are examined, and in the relativistic case the
dynamics of both massive and massless particles are investigated. The halo field
is given in general by a multipole expansion; we restrict ourselves to multipole
fields of pure order, whose Newtonian potentials are homogeneous polynomials
in cartesian coordinates. A pure \( n \)-pole field has \( n \) different
escapes, one of which is chosen by the particle according to its initial conditions.
We find that the escape has a fractal dependency on the initial conditions for
\( n>2 \) both in the Newtonian and the relativistic cases for massive test
particles, but with important differences between them. The relativistic motion
of massless particles, however, was found to be regular for all the fields we
could study. The box-counting dimension was used in each case to quantify the
sensitivity to initial conditions which arises from the fractality of the escape
route.
\end{abstract}

\section{Introduction}

It is common in astrophysics to find systems composed by an external material
halo surrounding an inner region which is either void or has a massive black
hole in the center. Examples of systems that are realistically described by
such a model are elliptical galaxies and globular clusters\cite{tremaine}.
The gravitational field in the inner region due to the halo satisfies the vacuum
field equations, and in Newtonian gravitation it can be expanded in a sum of
multipole terms of various orders. Due to the intrinsic nonlinearity of general
relativity, in general this is not true for general relativistic fields. However,
in many situations the halo can be supposed to be axisymmetric; in this case,
the general solution of Einstein's equations is known, and the inner relativistic
field can also be expanded in a series of multipolar terms. In this article,
we study the dynamics of test particles in pure multipolar halo fields (without
any matter in the inner region), both in the Newtonian and relativistic theories,
for several multipole orders, for material particles and (in the relativistic
cases) for light. We find that in each case the systems have distinct well-defined
\emph{escapes.} Different escapes are chosen by the test particles depending
on its initial conditions; if the escape route is a fractal function of the
initial conditions, the outcome of a particle starting from a given initial
condition shows a sensitive dependence on these initial conditions. In this
sense, we say that the system is chaotic\cite{grebogi,bleher,ott}. Associated
with this, we have a conveniently defined fractal dimension, which quantifies
this sensitiveness. Fractal escapes have been found in a number of physical
systems, including the Henon-Heiles system\cite{art2}, astrophysical potentials\cite{contop}, the motion of test particles
in gravitational waves\cite{gwav}, classical ionization\cite{ioniz}, motion
of particles in a fluid\cite{janosi}, inflationary cosmology\cite{cornish},
and others. We study the fractal dimension for different multipole orders and
for different energies, in the case of material test particles. For simplicity,
we confine ourselves to the case of zero angular momentum, when the motion is
planar.

This article is organized as follows: in section 2, we investigate the classical
multipole gravitational field, showing the fractality of the escape basins and
calculating the corresponding dimension for several multipole orders. In section
3, we turn to the relativistic case, and investigate the dynamics of both material
particles and massless particles; we compare the results with those obtained
in the classical case. In section 4, we summarize our results and draw some
conclusions.

\section{Newtonian multipole fields}

The Newtonian gravitational field in the inner region due to the mass in the
halo satisfies the vacuum (Laplace) field equation \( \nabla ^{2}\Psi =0 \),
and therefore can be uniquely expanded in a sum of multipole terms. If the field
is axisymmetric, as we admit from now on, its multipole expansion can be written
as:

\[
\Psi =\sum ^{\infty }_{n=0}\left\{ a_{n}R^{n}+b_{n}R^{-(n+1)}\right\} P_{n}(\cos \theta ),\]
 where \( R^{2}=z^{2}+r^{2} \), \( \cos \theta =z/R \), and \( r \) and \( z \)
are the usual cylindrical coordinates; the \( P_{n} \) are Legendre polynomials.
If we consider the inner region to be empty of matter, \( \Psi  \) cannot have
singularities there. This means that \( b_{n}=0 \) for all \( n \) in the
inner region. Thus:

\begin{equation}
\label{classexp}
\Psi =a_{0}+\sum ^{\infty }_{n=1}a_{n}R^{n}P_{n}(\cos \theta ),
\end{equation}
where \( a_{0} \) is a constant which can without lack of generality be chosen
as zero.

The multipole expansion of \( \Psi  \) contains only terms that become singular
as \( R\rightarrow \infty  \) (with the exception of the constant term \( a_{0} \)).
Here we are interested in studying the dynamics of the field due to each multipole
term separately. We therefore define the \( n \)-th order multipole field \( \Psi _{n} \)
as:

\begin{equation}
\label{psinclass}
\Psi _{n}=a_{n}R^{n}P_{n}(\cos \theta ).
\end{equation}

We now consider a test particle moving in the field \( \Psi _{n} \) with no
angular momentum in the direction of the symmetry axis \( z \); in this case,
the particle's motion is restricted to a plane that contains the \( z \) axis.
A convenient set of coordinates in this plane is given by the \( z \) axis
and the extended \( r \) axis, which reaches negative values. In order to avoid
confusion, we shall denote the extended \( r \) axis by \( x \), with \( -\infty <x<+\infty  \).
\( \Psi  \) is written in terms of \( x \) simply by replacing it for \( r \).
The Hamiltonian of a test particle of unit mass acted upon by the field \( \Psi _{n} \)
is:

\begin{equation}
\label{hamclass}
H_{n}=\frac{1}{2}\left( \dot{x}^{2}+\dot{z}^{2}\right) +\Psi _{n}(x,z).
\end{equation}

The corresponding equations of motion are \( \ddot{x}=-\partial \Psi _{n}/\partial x \),
\( \ddot{z}=-\partial \Psi _{n}/\partial z \). Since \( \Psi _{n} \) is a
homogeneous potential, satisfying \( \Psi _{n}(\lambda x,\lambda z)=\lambda ^{n}\Psi _{n}(x,z) \),
these equations are invariant with respect to the similarity transformations
\( x\rightarrow \lambda x \), \( z\rightarrow \lambda z \), \( t\rightarrow t/\sqrt{\lambda } \),
and \( E\rightarrow \lambda ^{n}E \), where \( \lambda  \) is an arbitrary
scale factor. Thus, to each orbit in the \( \Psi _{n}(x,z) \) potential there
corresponds a similar orbit in the transformed potential \( \Psi _{n}(\lambda x,\lambda z) \),
as long as \( \lambda  \) is positive. This means that the dynamics for all
positive values of \( E \) is the same except for scaling; we have to consider
only one value of \( E \) for each multipole potential \( \Psi _{n} \) when
we study its dynamical properties. We could use this scaling property of the
Hamiltonian to make \( a_{n}=1 \) in eq. (\ref{psinclass}), but we prefer
to keep the constant arbitrary because of numerical convenience.

The region in the \( (x,z) \)-plane accessible by a particle moving in the
potential \( \Psi _{n} \) with energy \( E \) is bounded by the equipotential
curves \( \Psi _{n}(x,z)=E \); the particle must satisfy \( \Psi _{n}(x,z)\leq E \).
This inequality can only be satisfied for a positive energy if \( \Psi _{n} \)
is positive. From equation (\ref{psinclass}), it follows that the directions
\( \theta  \) satisfying \( P_{n}(\cos \theta )<0 \) are accessible for all
positive energies, whereas directions satisfying \( P_{n}(\cos \theta )>0 \),
if followed far enough, reach values of \( R \) which violate the inequality
\( \Psi _{n}(x,z)\leq E \). Therefore, the directions with negative values
of \( P_{n}(\cos \theta ) \) define distinct \emph{escapes} of the potential
\( \Psi _{n} \); these escapes are separated by inaccessible regions lying
at angles with \( P_{n}(\cos \theta )>0 \). From the properties of Legendre
polynomials, it follows that the potential \( \Psi _{n} \) has \( n \) distinct
escapes.

As an example, Fig. 1 shows equipotential curves for some energies for the octopole
field \( \Psi _{3}. \) The curves for different energies have the same shapes,
which is a direct consequence of the potential's homogeneity. The three openings
in the equipotential curves are seen clearly.

To each escape we define the corresponding \emph{basin}, which is the set of
initial conditions in the phase space that lead to that escape. If the escape
chosen by a particle is a fractal function of its initial condition, then the
basins of the various escapes are mixed in a complex fashion, down to arbitrarily
small scales, and the final state of the particle (i.e. which escape it chooses)
can be extremely sensitive to its initial state. This happens when the basin
boundary is fractal. We then say that the system presents chaos. This is the
appropriate definition of chaos to open systems, whose motions are not confined
to a limited volume of the phase space, and this is the definition used throughout
this article.

The directions \( \theta _{e} \) which are minima of \( P_{n}(\cos \theta ) \)
correspond to valleys in the potential \( \Psi _{n} \); the escape regions
are centered around these directions. The walls on either side of these valleys
became increasingly steeper as we go to \( R\rightarrow \infty  \). This makes
the orbits of escaping particles become focused in beams around these directions
as they move towards \( R\rightarrow \infty  \). This phenomenon can be seen
in fig. 2a, which shows some numerically integrated orbits all beginning with
initial conditions on \( x=z=0 \), and varying initial velocity direction angle
\( \phi  \), defined by:
\[
v_{x}=v\sin \phi ;\quad v_{z}=v\cos \phi ,\]
 where \( v_{x} \) and \( v_{z} \) are the initial velocities in the \( x \)
and \( z \) directions, and the velocity modulus \( v \) is fixed by the conservation
of energy:
\[
v=\sqrt{2E},\]
since \( \Psi _{n}(0,0)=0 \).

The fractal nature of the basin boundary can be observed in a plot of the escape
angle \( \theta _{e} \) as a function on the initial velocity direction angle.
This is shown in Fig. 3a. We had to choose a ``cutoff distance'' \( R_{0} \)
such that the integration stops when \( R \) reaches \( R_{0} \). In the case
of Fig.3, we chose \( R_{0}=10.0 \). The discreteness of the values assumed
by the escape angle is clearly seen in Fig. 3; this is an illustration of the
focusing effect. The magnification of a tiny region of fig. 3a shown in fig.
3b gives us a convincing proof that the basin boundary is fractal. An example
of the sensitivity to initial conditions implied by the fractality of the basin
boundary is given in fig. 2b, where three orbits with \( \phi  \) differing
by less that one thousandth of a radian choose different escapes. We see that
initially the orbits are indistinguishable, and after a time they separate rather
abruptly and go to their distinct escapes. Figures 2 and 3 were plotted with
\( E=1 \) and \( a_{3}=1 \), but as we explained above, the result is exactly
the same regardless of the parameters' values. We note that we have found no
bounded orbits for any of the \( \Psi _{n} \) potentials, which might have
been expected from the shapes of the equipotentials.

The reason why the escape basins are fractally mixed is that the particle can
bounce many times off the potential ``walls'' before escaping. This means
that associated with the fractality in the basin boundary we have also fractality
in the escape time. To see that this is so, we plot the time it takes for a
particle to escape (without discriminating among the various escapes) as a function
of its initial velocity angle \( \phi  \), for a fixed initial position \( x_{0} \),
\( z_{0} \). Figure 4 shows the results for the octopole potential, with \( x_{0}=z_{0}=0 \).
The magnification shown in fig. 4b shows clearly the fractal nature of the escape
time. Notice that fractal and non-fractal areas are mixed for all scales. The
fractal nature of the escape time is a result of the existence of a fractal
set of singularities in the escape time function, where the escape time diverges.
This set of singularities results from the intersection between the set of initial
conditions (a topological circle in this case) and the set in the phase space
composed of orbits that never escape for \( t\rightarrow \infty  \), which
in its turn is the stable manifold of the so-called \emph{invariant set}, which
is also fractal in our case and is composed of orbits that never escape for
both \( t\rightarrow \infty  \) and \( t\rightarrow -\infty  \).

The sensitivity to initial conditions implied by the fractal nature of the basin
boundary is made precise with the introduction of the \emph{box-counting dimension
\( d \)}, calculated in the following way. For a given initial position (which
is \( x_{0}=y_{0}=0 \) in our case), we pick a large number of random initial
velocity angles \( \phi  \), with \( 0<\phi <2\pi  \). For each of these we
integrate the equations of motion and see to which basin it belongs. We then
take two neighboring angles \( \phi -\epsilon  \) and \( \phi +\epsilon  \)
(with the same initial position), with \( \epsilon  \) being a small number,
and find to which basin they belong. If all these three initial conditions do
not belong to the same basin (if they lead to different escapes), then \( \phi  \)
is said to be an ``uncertain'' angle. The number \( n \) of uncertain angles
found in a sample of \( N \) randomly chosen points is a function of the displacement
\( \epsilon  \), and for large \( N \) the fraction \( n/N \) of uncertain
points scales as
\begin{equation}
\label{feps}
\frac{n}{N}\equiv f\propto e^{1-d},
\end{equation}
 where \( d \) is by definition the box-counting dimension of the intersection
of the basin boundary with the one-dimensional manifold in which we chose the
initial conditions. We numerically calculate \( f \) for several values of
\( \epsilon  \), and \( d \) is directly obtained from the plot of \( \ln f \)
versus \( \ln \epsilon  \), which should be a straight line, and whose inclination
\( \alpha  \) is related to \( d \) by \( \alpha =1-d \).

We have calculated in this way the box-counting dimension for multipole fields
with orders from 2 to 9. Since the dynamics is the same for every value of the
energy, \( d \) does not depend on \( E \), and is unique for each multipole
order. Figure 5 shows the plot of \( \ln f \) versus \( \ln \epsilon  \) we
have obtained for the octopole field (\( n=3 \)). The data fit quite well into
a straight line, whose inclination (\( \alpha =0.77\pm 0.01 \)) gives a dimension
\( d=0.23\pm 0.01 \), confirming the fractal nature of the basin boundary.
Figure 6 shows \( d \) as a function of the multipole order \( n \). We see
that \( d \) increases monotonically with \( n \), which means that the fractality
of the basin boundary increases with \( n \), and thus the dynamics becomes
``more chaotic'' with multipolar fields of higher orders. The dipole and quadrupole
fields (\( n=1 \) and \( n=2 \), respectively) have regular basin boundaries,
and show no chaos in their dynamics. In fact, the potentials \( \Psi _{1} \)
and \( \Psi _{2} \) are separated using the \( x,z \) coordinates, as can
be immediately seen from (\ref{psinclass}), which implies that the Hamiltonian
is integrable and the basin boundaries are regular.

A field of the general form (\ref{classexp}), with the contribution of more
than one multipole order, is not homogeneous, and its dynamics changes with
the energy. We have studied how the box-counting dimension changes with \( E \)
for the field \( \Psi =\Psi _{3}+\Psi _{4} \), with equal contributions of
the 3rd and 4th multipole orders. For high values of \( E \), the particle's
motion is dominated by the 16-pole component of the field, and the equipotentials
are practically the same as the pure 16-pole field. Accordingly, for these energies
the box-counting dimension was found to be numerically indistinguishable from
the value we obtained for the pure 16-pole field \( \Psi _{4} \). As the energy
is lowered, however, \( d \) does \emph{not} go to its pure octopole field
value, as might be imagined. In fact, for \( E=0.1 \), we found \( d=0.38\pm 0.01 \);
this value of \( d \) is higher then that of both the pure octopole field and
the quadrupole field. Thus, the fractal dimension from a field made by the mixture
of two multipole orders is not necessarily an intermediate value between the
dimensions of each pure multipole order separately.

\section{Relativistic multipole fields}

In this section we investigate the dynamics with a general-relativistic generalization
of the Newtonian multipole fields we studied in the previous section. In general
relativity, the dynamics results from the space-time metric \( ds^{2}=g_{\mu \nu }dx^{\mu }dx^{\nu } \),
where \( g_{\mu \nu } \) is the metric tensor and \( x^{\mu } \) are the space-time
coordinates. The equations of motion for a test particle in the field \( g_{\mu \nu } \)
are given by the geodesic equation \( \ddot{x}^{\mu }=\Gamma ^{\mu }_{\alpha \beta }\dot{x}^{\alpha }\dot{x}^{\beta }, \)
where \( \Gamma ^{\mu }_{\alpha \beta } \) is the Christoffel symbol, and \( \dot{x}^{\mu }=dx^{\mu }/d\lambda  \),
with \( \lambda  \) being a convenient affine parameter.

For an axially symmetric and static field, the metric is most conveniently written
in the Weyl form\cite{kramer}:

\begin{equation}
\label{weyl}
ds^{2}=e^{2\Psi }dt^{2}-e^{-2\Psi }\left[ e^{2\gamma }\left( dr^{2}+dz^{2}\right) +r^{2}d\phi ^{2}\right] ,
\end{equation}
where \( r \), \( z \) and \( \phi  \) are cylindrical-like coordinates,
and \( t \) is the time. We use units such that \( c=1 \) throughout. The
metric functions \( \Psi  \) and \( \gamma  \) depend on \( r \) and \( z \)
only, and satisfy the vacuum Einstein field equations: 

\begin{equation}
\label{eq1}
\nabla ^{2}\psi \equiv \frac{\partial ^{2}\psi }{\partial r^{2}}+\frac{1}{r}\frac{\partial \psi }{\partial r}+\frac{\partial ^{2}\psi }{\partial z^{2}}=0;
\end{equation}
\begin{equation}
\label{eq2}
d\gamma =r\left[ \left( \frac{\partial \psi }{\partial r}\right) ^{2}-\left( \frac{\partial \psi }{\partial z}\right) ^{2}\right] dr+2r\frac{\partial \psi }{\partial r}\frac{\partial \psi }{\partial z}dz.
\end{equation}
 Eq. (\ref{eq1}) is just Laplace's equation in cylindrical coordinates, and
eq. (\ref{eq2}) is a quadrature whose integrability is automatically guaranteed
by eq. (\ref{eq1}). Thus, each axisymmetric solution of Newton's field equation
corresponds to a general-relativistic metric (\ref{weyl}), with \( \gamma  \)
being found by integration of eq. (\ref{eq2}) with the requirement that the
metric be regular in the symmetry axis. Based on this, we define the relativistic
equivalent to the multipolar fields \( \Psi _{n} \) in eq. (\ref{psinclass})
to be the metric (\ref{weyl}) with \( \Psi =\Psi _{n} \) and \( \gamma =\gamma _{n} \)
obtained through eq. (\ref{eq2}). The integration can be performed analytically,
and the result is:
\begin{equation}
\label{gamman}
\gamma _{n}=\frac{n}{2}a^{2}R^{2n}\left[ P_{n}^{2}(\cos \theta )-P_{n-1}^{2}(\cos \theta )\right] ,
\end{equation}
with \( R^{2}=r^{2}+z^{2} \). Since there are many possible relativistic space-times
which in the weak field limit correspond to a given Newtonian field, the choice
of the metric is not unique. We have chosen the metric defined by \( \Psi _{n} \)
and \( \gamma _{n} \) because it is the simplest choice, and because the field
obtained in this way shares important features with the Newtonian field, as
we shall see shortly.

As in the Newtonian case, the invariance of the system with respect to time
translations and axial rotations implies the existence of two constants of motion,
which can be identified as the energy \( E \) and the \( z \)-component of
the angular momentum \( L_{z} \). They can be found by using the Lagrangian
associated with the metric (\ref{weyl}):
\begin{equation}
\label{lagr}
2L=e^{2\Psi }\dot{t}^{2}-e^{-2\Psi }\left[ e^{2\gamma }\left( \dot{r}^{2}+\dot{z}^{2}\right) +r^{2}\dot{\phi }^{2}\right] ,
\end{equation}
where the dot denotes differentiation with respect to the affine parameter.
Since both \( \Psi  \) and \( \gamma  \) depend only on \( r \) and \( z \),
the Lagrangian \( L \) does not depend on \( t \) and \( \phi  \); they are
``cyclical coordinates'' in the language of Hamiltonian dynamics, and they
are associated with the conserved quantities:

\begin{equation}
\label{relE}
E=\frac{\partial L}{\partial \dot{t}}=e^{2\Psi }\dot{t};
\end{equation}
\begin{equation}
\label{relL}
L_{z}=\frac{\partial L}{\partial \dot{\phi }}=-r^{2}e^{-2\Psi }\dot{\phi }.
\end{equation}

Since the Lagrangian function (\ref{lagr}) has only kinetic terms, and since
\( L \) does not depend explicitly on the affine parameter, the Hamiltonian
of this system is simply \( L \) itself, and its conservation gives us yet
another constant of motion:
\[
L=\frac{1}{2}m_{0}^{2},\]
where \( m_{0} \) is the test particle's rest mass. In the case of light, we
have \( m_{0}=0 \); for massive particles, we can without loss of generality
make \( m_{0}=1 \). Rewriting \( L \) with the help of eqs. (\ref{relE})
and (\ref{relL}), we have:
\begin{equation}
\label{hamrel}
E^{2}e^{-2\Psi }-L_{z}^{2}r^{-2}e^{2\Psi }-e^{2(\gamma -\Psi )}\left( \dot{r}^{2}+\dot{z}^{2}\right) =\delta ,
\end{equation}
with \( \delta =0 \) for particles with zero rest mass, and \( \delta =1 \)
for massive particles. As in the Newtonian case in the previous section, we
will restrict ourselves to motion with zero angular momentum (\( L_{z}=0 \)),
when the motion is planar. In complete analogy to what we did for the Newtonian
potential, we define a new coordinate \( x \), which is no more than \( r \)
extended to negative values; the motion of our zero angular momentum test particle
occurs on the \( (x,z) \)-plane. All the expressions above are written in terms
of \( x \) simply by replacing \( r \) by \( x \) everywhere. 

The accessible region of the \( (x,z) \)-plane is found by setting \( \dot{x}=\dot{z}=0 \)
in (\ref{hamrel}); taking into account that \( L_{z}=0 \), we have:
\begin{equation}
\label{potefrel}
E^{2}(\dot{x}=\dot{z}=0)\equiv U(x,z)=\delta e^{2\Psi }.
\end{equation}

The accessible region for a particle with energy \( E \) is given by \( U(x,z)\le E^{2} \).
For massless particles (\( \delta =0 \)), \( U=0 \), and this condition is
always satisfied; thus, the whole \( (x,z) \)-plane is accessible to null geodesics.
This is not so for massive particles (\( \delta =1 \)); the regions they are
able to reach on the \( xz \) plane depend on their energy and are determined
by \( \Psi  \).

We now consider the pure multipole potentials \( \Psi _{n}=a_{n}R^{n}P_{n}(\cos \theta ) \),
with \( \gamma _{n} \) given by (\ref{gamman}). We first treat the case of
massive particles; the motion of light will be studied later. The effective
potential is \( U=e^{2\Psi }=g_{tt} \), where \( g_{tt} \) is the time-time
component of the metric. It is clear that the dynamics in the relativistic field
\( (\Psi _{n},\gamma _{n}) \) is more complex than its Newtonian counterpart.
In particular, since neither the effective potential (\ref{potefrel}) nor \( \gamma _{n} \)
are homogeneous functions, and furthermore the relativistic equations of motion
depend on the velocities \( \dot{x}^{\mu } \), the orbits in the relativistic
field do not satisfy the homogeneity property that guarantees for the orbits
in the Newtonian field that the dynamics is the same for all energies, up to
a global scaling. We thus expect the dynamics to change with the energy in the
relativistic case; this is a fundamental difference between the two theories.
In spite of this, we see from eq. (\ref{potefrel}) that the curves of constant
effective potential are still given by \( \Psi _{n}=const \) (although the
constant is no longer just the energy). This means that the escapes in the potentials
\( \Psi _{n} \) are the same as those in the Newtonian fields. As \( R\rightarrow \infty  \)
along the asymptotically allowed directions, given by \( P_{n}(\cos \theta )<0 \)
(exactly as in the Newtonian case), we have \( U=g_{tt}\rightarrow 0 \), that
is, we have an infinite red shift (as in section 2, we assume \( a_{n}>0 \)):
particles are attracted to these regions. On the other hand, for asymptotically
forbidden regions, with \( P_{n}(\cos \theta )<0 \), we have \( g_{tt}\rightarrow \infty  \),
an infinite blue shift: particles are attracted \emph{away} from these regions.

We now study the dynamics in this field using the same techniques we used in
the Newtonian case. We integrate numerically orbits with initial position \( x_{0}=z_{0}=0 \)
and variable velocity angle \( \phi  \), with the magnitude \( v \) of the
velocity being fixed by the conservation of energy. Since \( \Psi _{n}(0,0)=\gamma _{n}(0,0)=0 \)
for all \( n \), from eq. (\ref{hamrel}) we see that:

\[
\dot{x}_{0}=v\sin \phi \; \; \dot{z}_{0}=v\cos \phi ,\]
with

\[
v^{2}=E^{2}-1.\]

The equations of motion for the independent variables \( x \) and \( z \)
are obtained directly from the geodesci equations:

\begin{equation}
\label{eqmot1}
\ddot{x}=-\frac{1}{2f}\left[ g^{tt}_{,x}E^{2}+f_{,x}\left( \dot{x}^{2}-\dot{z}^{2}\right) +2f_{,z}\dot{x}\dot{z}\right] ;
\end{equation}
\begin{equation}
\label{eqmot2}
\ddot{z}=-\frac{1}{2f}\left[ g^{tt}_{,z}E^{2}+f_{,z}\left( \dot{z}^{2}-\dot{x}^{2}\right) +2f_{,x}\dot{x}\dot{z}\right] ,
\end{equation}
with \( f=e^{2(\Psi -\gamma )} \) and \( g^{tt}=e^{-2\Psi } \); these equations
are valid for \( L_{z}=0 \). Together with the quadrature (\ref{relE}), the
above equations give the complete dynamics of the system.

Similar to what we found in the previous section, the orbits are focused along
directions corresponding to the minima of \( P_{n}(\cos \theta ) \). Plots
of the escape angle for \( n>2 \) are very similar to Fig. 3, showing clearly
that the escape has a fractal dependency on the initial conditions. As in the
Newtonian case, the dipole and quadrupole fields (\( n=1 \) and 2, respectively)
show a regular basin boundary. However, the equations of motion are not separated
in \( x,z \) coordinates, as is the case for the Newtonian fields.

As we explained above, we expect the fractal basin boundary present in the \( n>2 \)
fields to change with the energy in the relativistic case, as opposed to the
Newtonian one. As in the Newtonian case, we have found no bounded orbits, neither
for massive particles nor for massless ones.

We calculate the fractal dimension of this set in the same way we did previously
for the motion in the Newtonian field. In that case the dimension did not depend
on the energy, because the dynamics is the same for all (positive) energies.
This is not true for the relativistic dynamics, and we expect the dimension
to change with the energy. This is indeed what we have found. As an example,
we take the field \( (\Psi _{4},\gamma _{4}) \). We consider particles with
energies higher than 1, corresponding to positive energies in the Newtonian
case. For \( E \) close to 1 (corresponding to energies close to 0 in the Newtonian
field), we find that the box-counting dimension is equal to its Newtonian value.
For instance, for \( E=1.01 \), we find \( d=0.33\pm 0.02 \), which is (within
the error) the same value we found for the Newtonian field \( \Psi _{4} \).
For higher energies, however, \( d \) changes with \( E \): for \( E=1.5 \),
we have \( d=0.30\pm 0.02 \), and for \( E=2.0 \), \( d=0.26\pm 0.02 \).
We have calculated \( d \) for more values of \( E \) than we show here, and
it seems that \( d \) decreases monotonically with \( E \), and the system
becomes more chaotic for large energies; this is the contrary to what is usually
the case in bounded dynamics, where commonly the chaos grows with the energy.
The same behavior is observed in the octopole field (\( n=3 \)), but in this
case the variations in the dimension are smaller. Because of the complexity
of the equations of motion for high multipole orders, we have not been able
to calculate \( d \) with enough accuracy for \( n \) larger than 4, but we
think this behavior of \( d \) might be a general feature of the dynamics for
the \( \Psi _{n} \) fields. 

In the case of null geodesics, the equations of motion are again given by eqs.
(\ref{eqmot1}) and (\ref{eqmot2}), but the initial conditions are required
to satisfy the constraint (\ref{hamrel}) with \( \delta =0 \) (and \( L_{z}=0 \)
in our case). This allows us to scale the affine parameter so as to eliminate
\( E \) from the equations. This means that the planar motion of light depends
only on the chosen field \( \Psi _{n} \), and not on any other parameter, unlike
the motion of massive particles. 

Even though the whole \( (x,z) \)-plane is accessible to massless particles,
as we have seen, null geodesics also show focusing effects similar to that shown
by massive particles. An example is given in Fig. 7, where the escape angle
is plotted as a function of the initial velocity angle for null geodesics, for
the octopole field \( n=3 \). We see clearly that the values assumed by the
escape angle are indeed discrete. We do not see, however, any region where the
escape angle shows rapid variations, that is typical of fractal functions. Magnifications
of several regions of Fig. 7 show no small-scale structures, indicating that
the escape angle is not fractal. This is confirmed by direct calculation of
the box-counting dimension, which gives \( d=1 \) to within statistical accuracy.
This result also holds for \( n=4 \) and \( n=5 \), which is the highest multipole
order for which we have been able to obtain reliable results. We speculate that
this result might be true for all \( n \).

\section{Conclusions}

We have studied the dynamics of test particles in pure multipole fields generated
by external halos of matter, when the particles have zero angular momentum and
the motion happens on a plane. Such a system could be used to approximate observed
astronomical fields, such as the interior of galaxies. We have considered both
the Newtonian and the general-relativistic fields, and we have compared the
results of the two theories. 

For the Newtonian fields, the homogeneity of the pure potentials \( \Psi _{n} \)
implies that the dynamics is independent on the energy. We have found that the
escape chosen by the test particles is a fractal function of the initial conditions
for all pure multipole fields of order higher than 3. For \( n=1,2 \), the
Hamiltonian is separable and the basin boundaries are regular. We have calculated
numerically the box-counting dimension \( d \) associated with the escape function.
Since the dynamics is the same for all energies, \( d \) depends only on the
multipole order \( n \). We have found that \( d \) is an increasing function
of \( n \) for \( n \) up to 9; higher values of \( n \) could not be studied
because of numerical limitations. The increasing in the intensity of chaos with
\( n \) can be understood by the fact that fields of higher order have a higher
number of escapes. We have also studied the dynamics of a field made by the
superposition of two different multipole orders, specifically, the field \( \Psi =\Psi _{3}+\Psi _{4} \).
For this field, the dynamics depends on the energy, and we found that, although
for high energies the box-counting dimension tends to the value of the \( \Psi _{4} \)
field, for low energies \( d \) assumes values higher than the dimensions of
both \( \Psi _{3} \) and \( \Psi _{4} \), maybe somewhat surprisingly.

We now discuss the results found for the relativistic fields, starting by the
dynamics of massive particles. The relativistic pure multipole fields have not
the homogeneous property shown by their Newtonian counterparts, and their dynamics
does depend on the energy, and so does the escape function. We found that the
escape chosen by the particle is a fractal function of the initial conditions
for fields of order higher than 2 also in the relativistic case. The dipole
and quadrupole fields have regular basin boundaries, but unlike the Newtonian
fields, the relativistic equations of motion are not directly separable. For
\( n>2 \), we expect the box-counting dimension \( d \) to change with the
energy. We have calculated \( d \) as a function of the energy, and we have
found that for low energies, \( d \) goes to the Newtonian values, as might
have been expected. As \( E \) increases, \( d \) was found to decrease monotonically,
for all the energies for which we have been able to obtain reliable results.
We have obtained these results for \( n=3,4 \) and 5, and this seems to be
a general feature of the dynamics of these fields. 

For massless test particles, the escape function was found to be regular (non-fractal)
for \( n \) up to 5; at least as far as the fifth multipole order, null geodesics
are not chaotic.

\section*{Acknowledgments}

We would like to thank FAPESP and CNPq for their financial support.

\newpage

\section*{Figure captions}

\begin{description}
\item [Fig.~1]Equipotential curves for the Newtonian octopole field \( \Psi _{3} \).
From the inside out, the energies are \( E=0.25 \), \( E=1.0 \) and \( E=4.0 \).
\item [Fig.~2]Orbits in the Newtonian field \( \Psi _{3} \) all starting from \( x=z=0 \),
with different initial directions and \( E=1.0 \), \( a_{3}=1 \). Fig. 2b
shows three orbits with initial velocity angles differing by 1/5000 radian;
the circle marks the starting point \( x=z=0 \).
\item [Fig.~3]Escape angle \( \theta _{e} \) as a function of the initial velocity
angle \( \phi  \) for the octopole field. Fig. 3b is a magnification of a small
region of Fig. 3a.
\item [Fig.~4]Escape time as a function of the initial velocity angle \( \phi  \)
for the octopole field. Fig. 4b is a magnification of a small region of Fig.
4a.
\item [Fig.~5]\( \ln f \) vs. \( \ln \epsilon  \) for the Newtonian octopole field.
The straight line is the result of the fitting.
\item [Fig.~6]Box-counting dimension \( d \) for Newtonian multipole field of order
\( n \). The error bars come from the statistical errors in the calculation
of \( d \).
\item [Fig.~7]Escape angle \( \theta _{e} \) as a function of the initial velocity
angle \( \phi  \) for the null geodesics in the relativistic octopole field. 
\end{description}
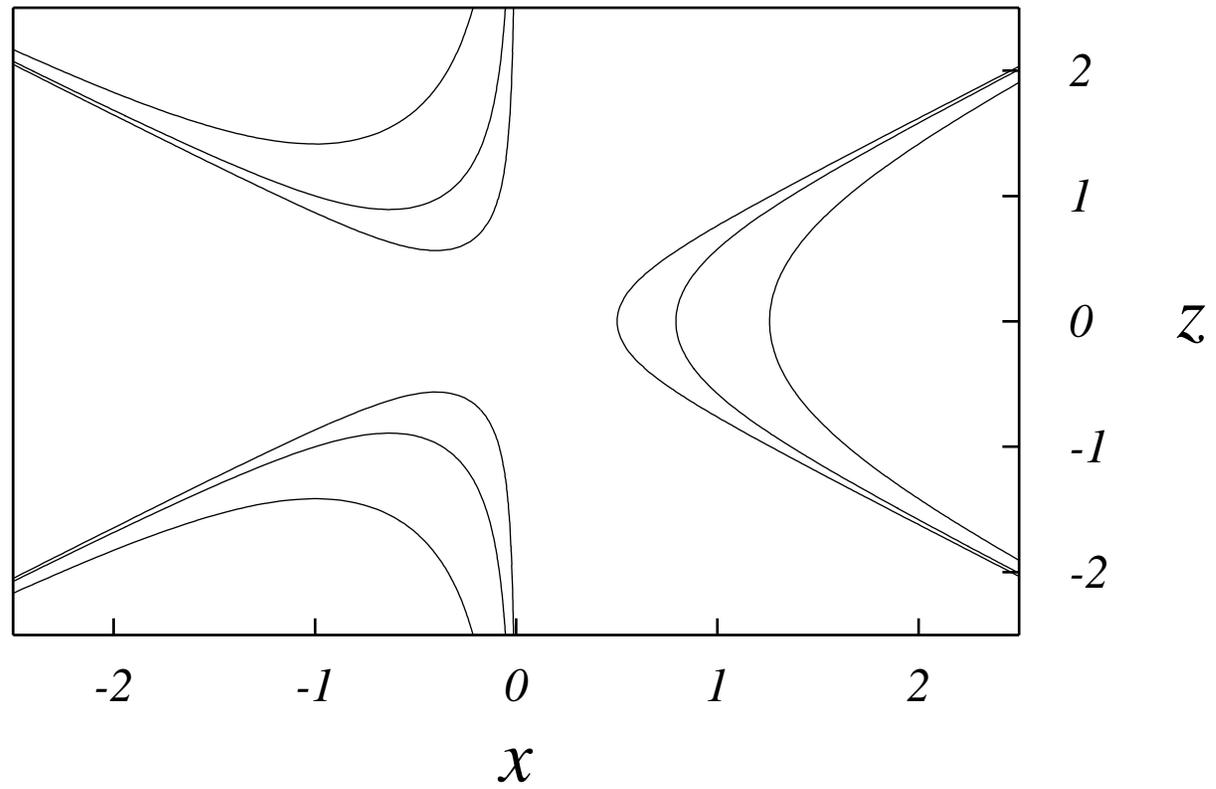
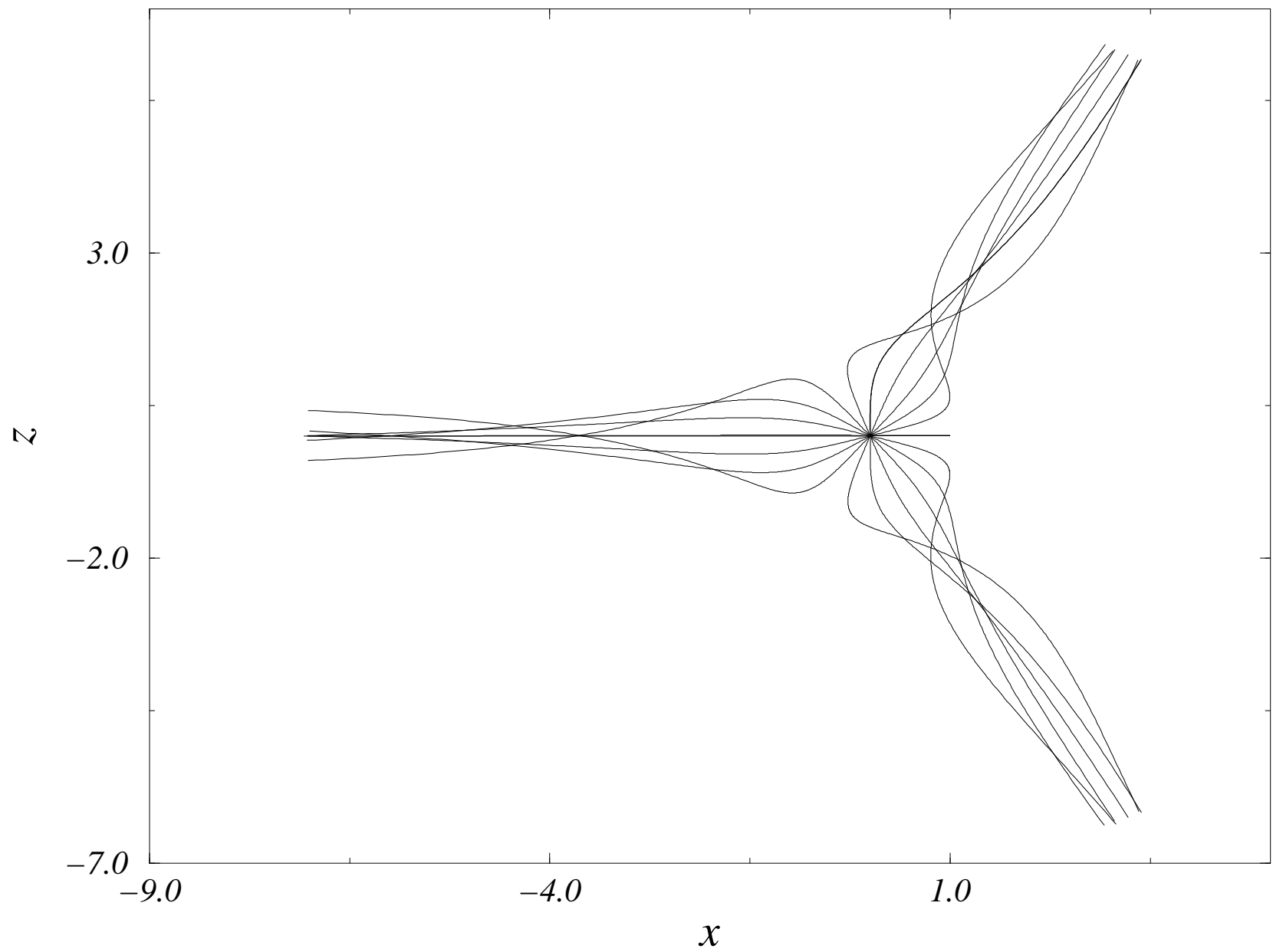
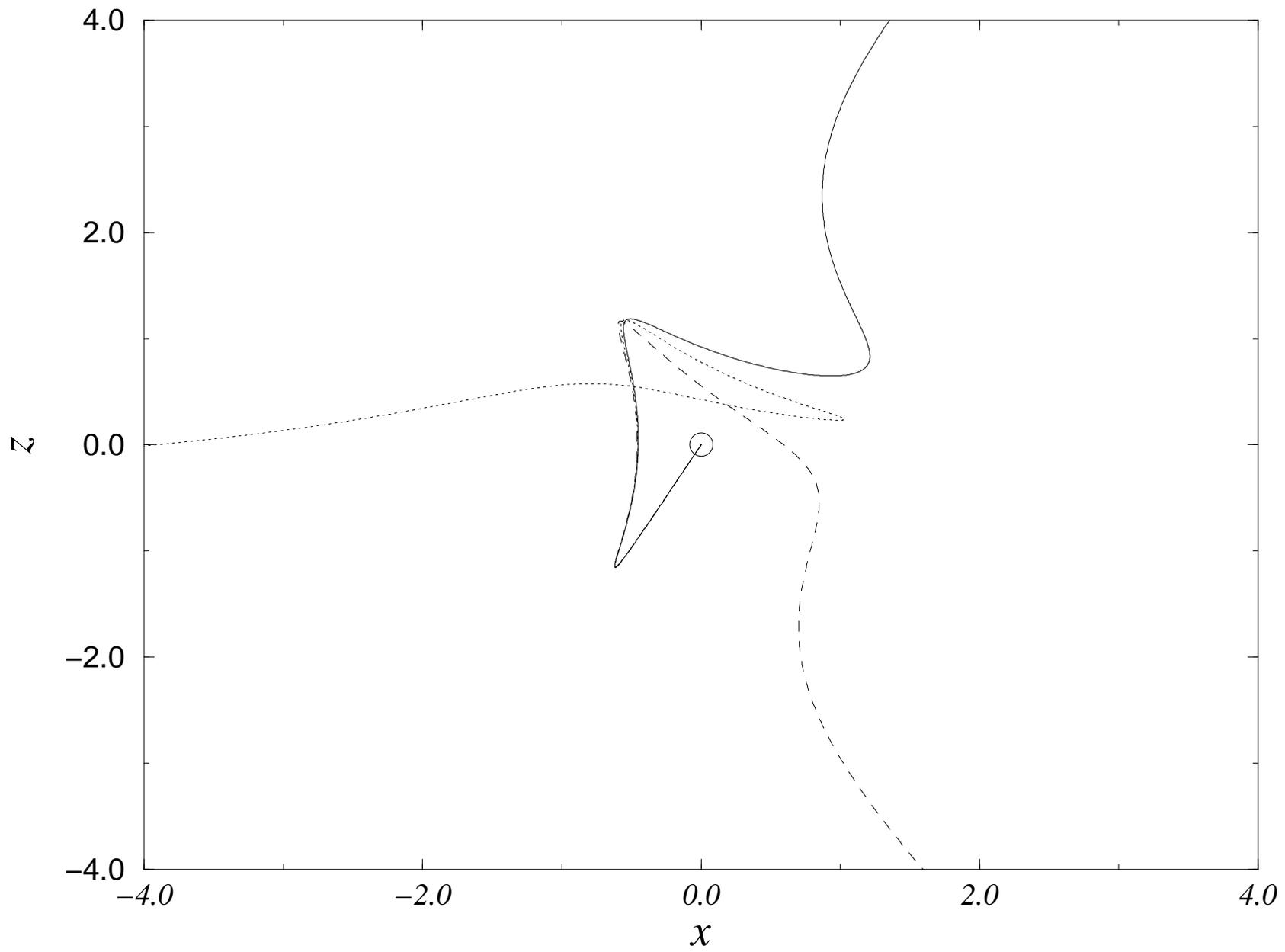
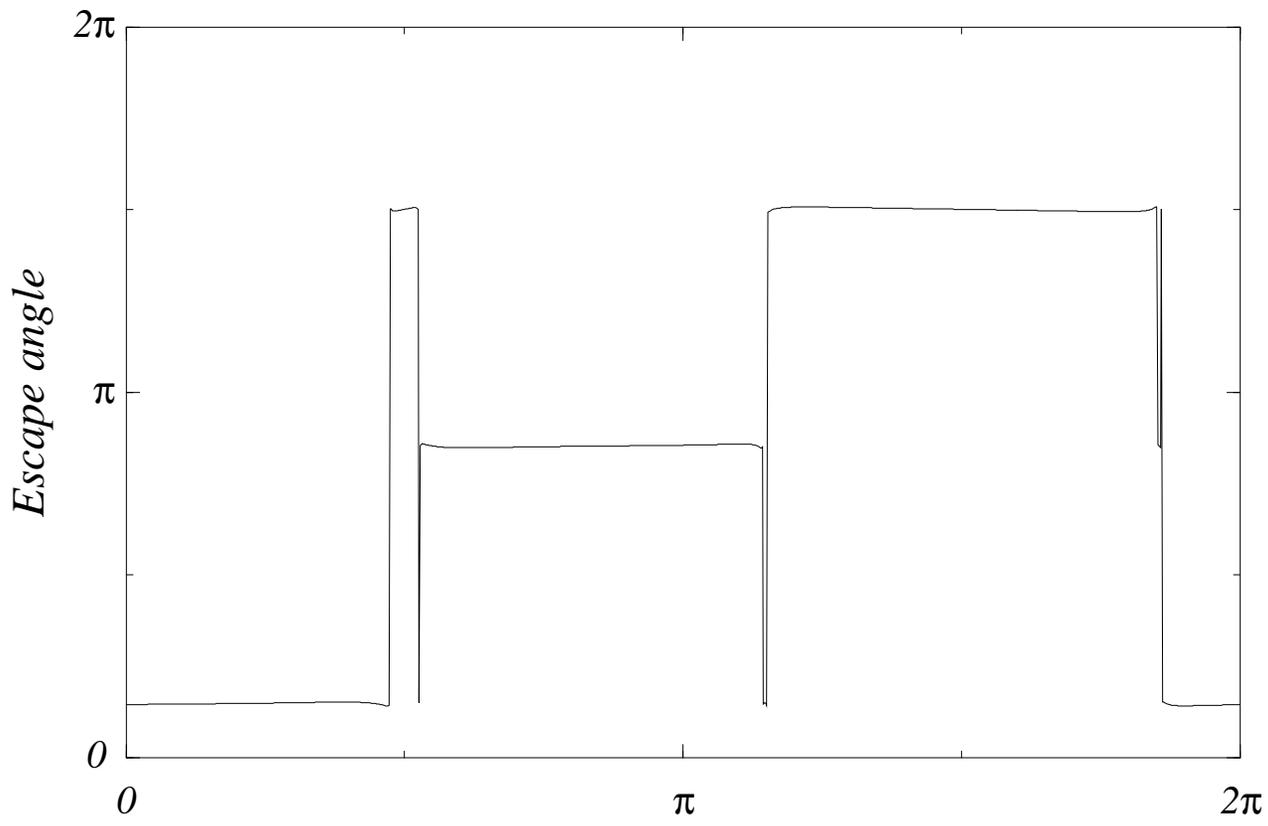
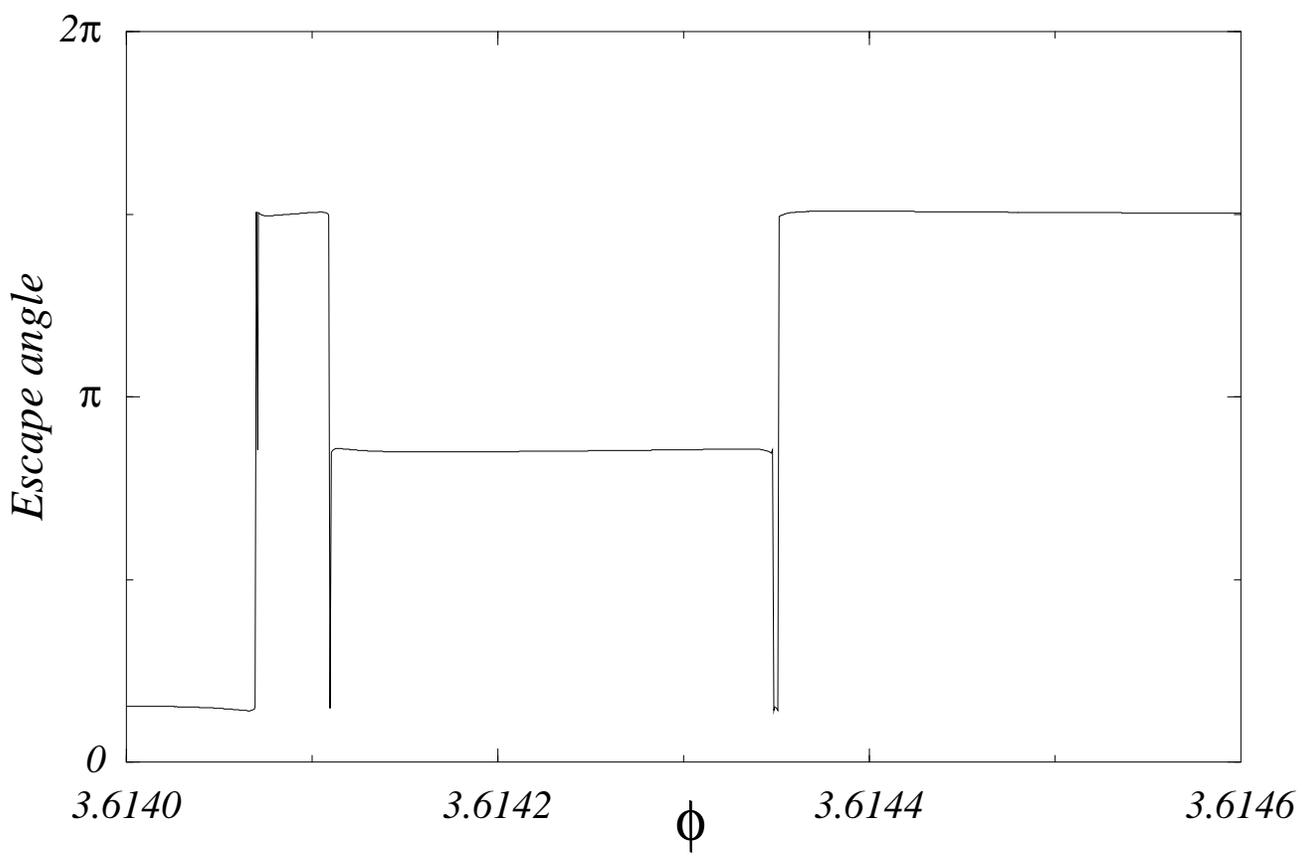
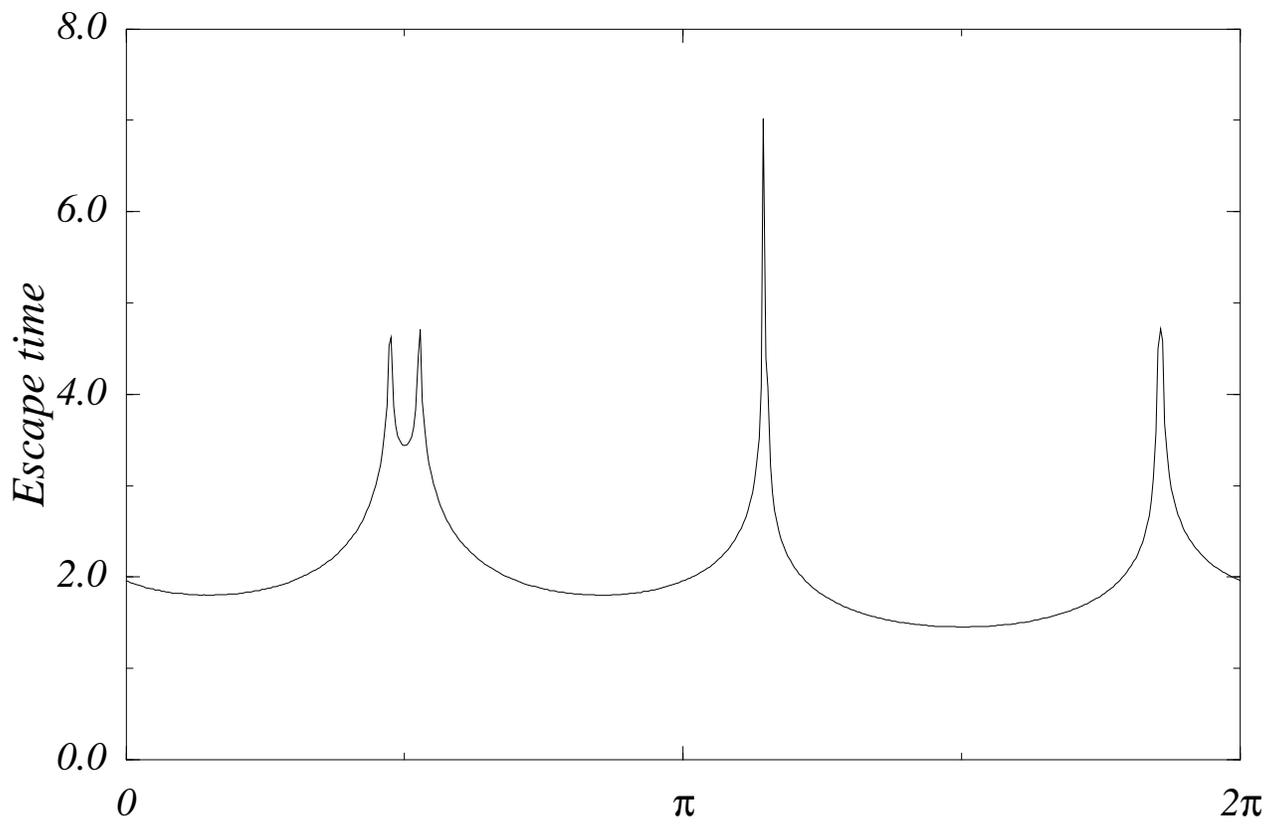
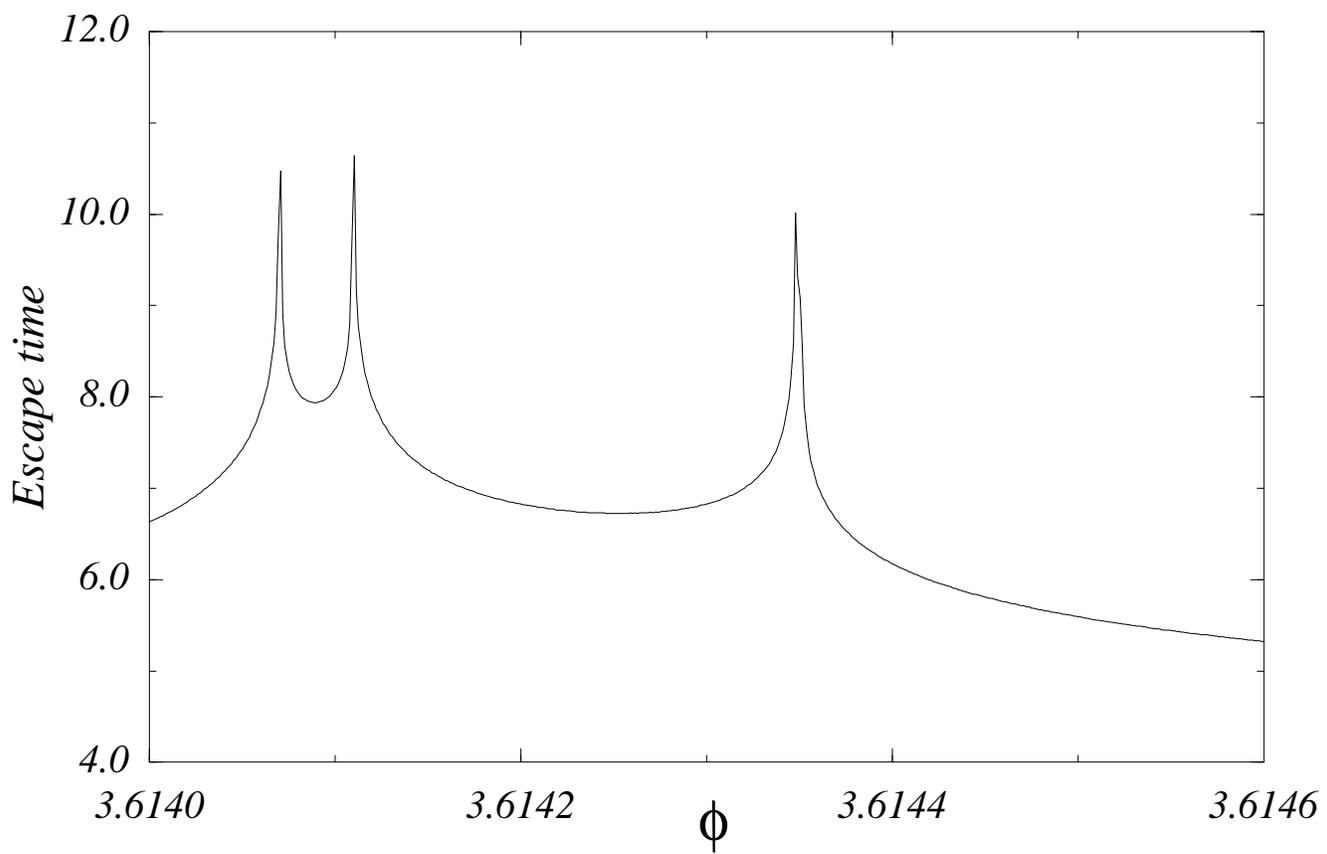
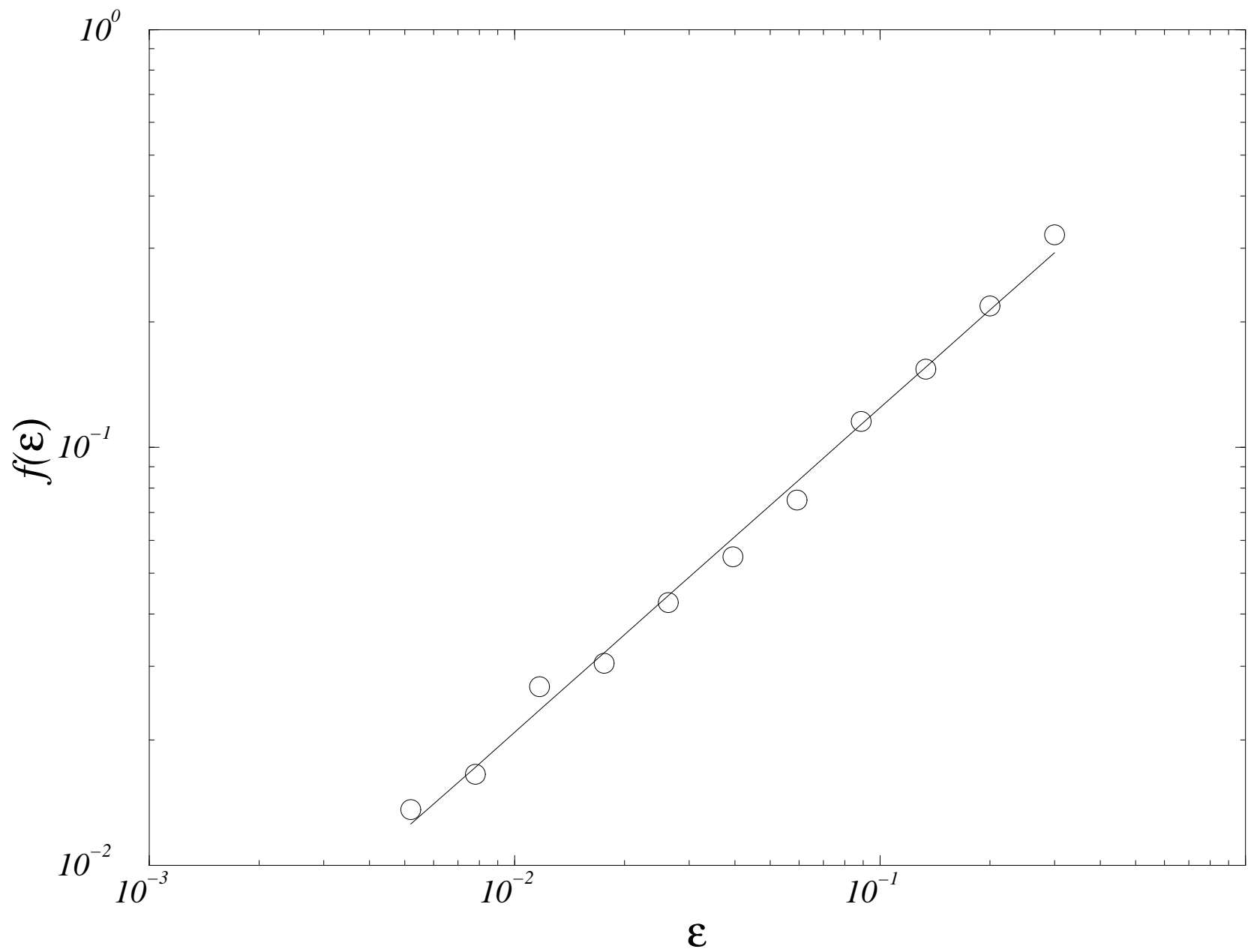
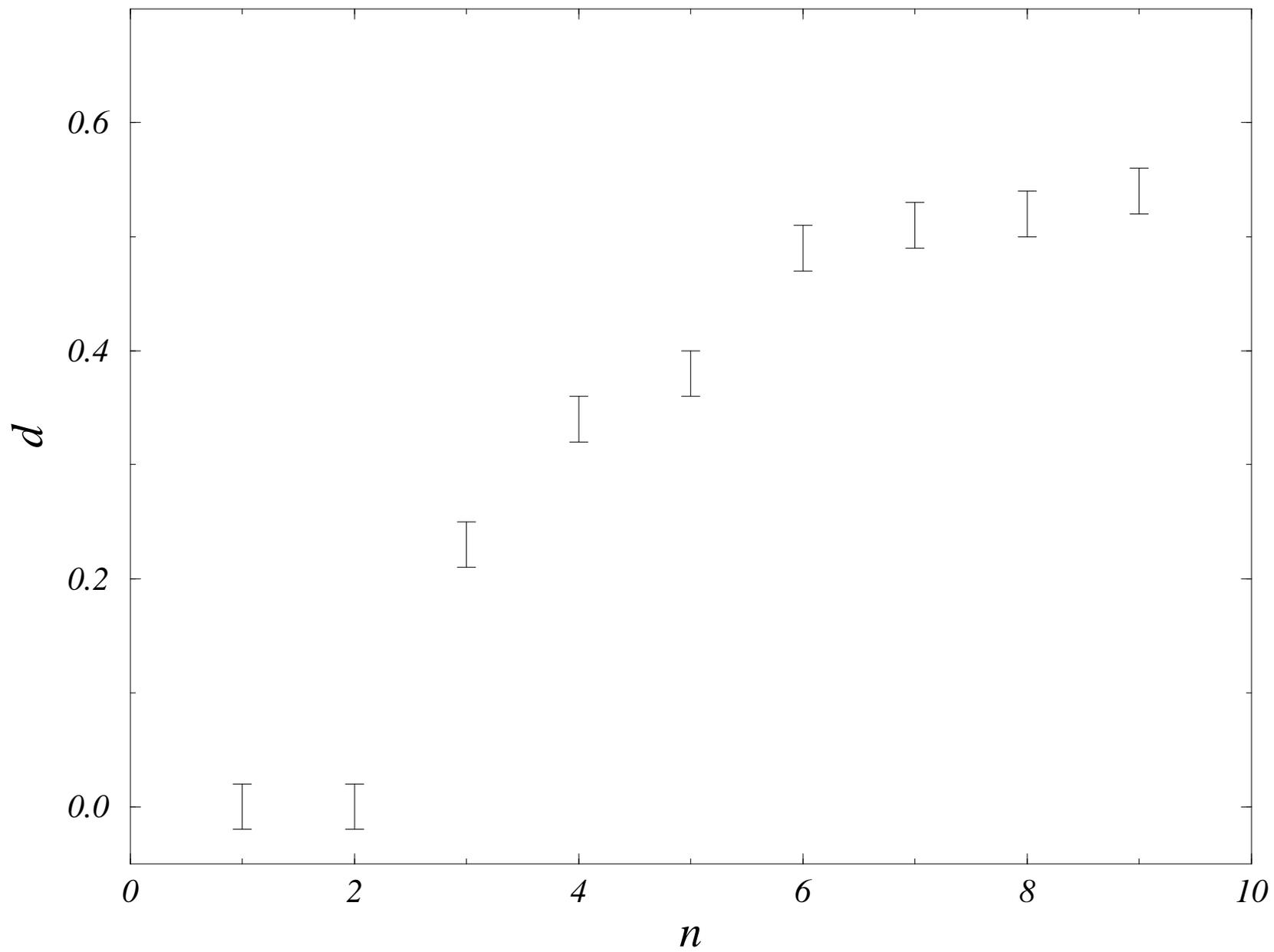
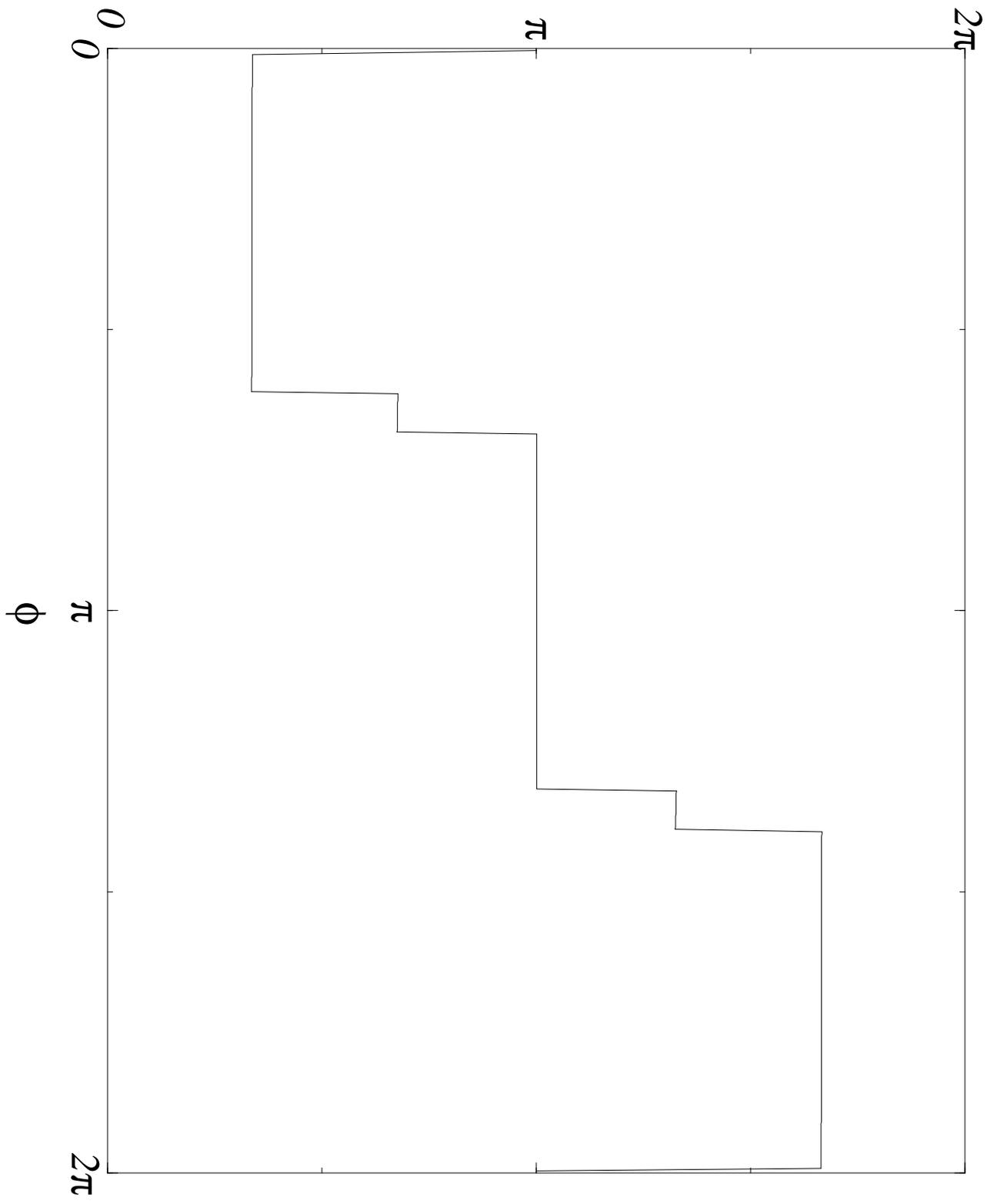
\end{document}